\documentclass[aps,prl,twocolumn,superscriptaddress,letterpaper]{revtex4}
\usepackage{graphicx}
\usepackage{siunitx}
\usepackage{color}
\usepackage{mdwlist}
\usepackage[version=3]{mhchem}

\DeclareSIUnit\elementarycharge{\text{\ensuremath{e}}}
\DeclareSIUnit\formulaunit{\text{\ensuremath{f.u.}}}
\DeclareSIUnit\bohrmagnetron{\text{\ensuremath{\mu_\mathrm{B}}}}
\DeclareSIUnit\percent{\text{\ensuremath{\%}}}

\begin{document}

\title{Designing a fully-compensated half-metallic ferrimagnet}

\author{Mario {\v Z}ic}
\affiliation{CRANN and School of Physics, Trinity College Dublin, Dublin 2, Ireland}
\author{Karsten Rode}
\author{Naganivetha Thiyagarajah}
\author{Yong-Chang Lau}
\author{Davide Betto}
\author{J.M.D. Coey}
\author{Stefano Sanvito}
\affiliation{CRANN, AMBER and School of Physics, Trinity College Dublin, Dublin 2, Ireland}
\author{Kerry J. O'Shea}
\author{Ciaran A. Ferguson}
\author{Donald A. MacLaren}
\affiliation{SUPA, School of Physics \& Astronomy, University of Glasgow, Glasgow G12 8QQ, United Kingdom}
\author{Thomas Archer}
\email[]{archert@tcd.ie}
\homepage[www.materials-mine.com]{}
\affiliation{CRANN and School of Physics, Trinity College Dublin, Dublin 2, Ireland}

\date{\today}

\begin{abstract}
  Recent experimental work on Mn$_2$Ru$_x$Ga demonstrates its potential as a
  compensated ferrimagnetic half-metal (CFHM).
  Here we present a set of high-throughput \emph{ab initio} density functional theory
  calculations and detailed experimental characterisation,
  that enable us to correctly describe the nominal Mn$_2$Ru$_x$Ga thin films,
  in particular with regard to site-disorder and defects.
  We then construct models that accurately capture all the key features of
  the Mn-Ru-Ga system, including magnetic compensation and
  the spin gap at the Fermi level.
  We find that electronic doping is neccessary,
  which is achieved with a Mn/Ga ratio smaller than two.
  Our study shows how composition and substrate-induced biaxial strain
  can be combined to design the first room-temperature CFHM.
\end{abstract}


\maketitle


Compensated ferrimagnetic half metals (CFHM) have an ordered spin-state which results in half-metallicity, without any net magnetic moment. 
As the material creates no stray magnetic field it should have low Gilbert damping, 
and offer numerous advantages compared to standard ferromagnetic metals. 
These include higher frequency operation, higher packing density, 
reduced device power requirement and devices that are impervious to external magnetic fields. 
Although there is no net magnetic moment, the highly-polarised spin-state 
allows switching of the magnetisation via spin-transfer torque. 
The class of CFHMs was first envisaged by van Leuken and de Groot\cite{leuken1995} in 1995, but despite significant effort\cite{coey2013,yang,hori,fesler2006,galanakis2006,prb2009} the goal of a CFHM had proved elusive.\cite{xiao2012}

Recent  experimental\cite{coey2014, coey2015, coey2015b} and theoretical\cite{blugel2014} efforts towards creating a CFHM have concentrated on the Heusler alloy system Mn$_2$Ru$_x$Ga (MRG).
Heusler alloys are made of four interlaced \emph{fcc} lattices, 
which form a \emph{bcc}-like structure.
Mn$_2$RuGa has the full Heusler (L2$_1$) structure, with Mn occupying the $4a$ and $4c$ sites, Ru the $4d$ and Ga the $4b$ sites.\cite{coey2014}
Half-metallic Heuslers with the L2$_1$ structure are expected to follow a modified Slater-Pauling curve 
with the net magnetic moment $m$ given by $m=N_v-24$ where $N_v$ is the number of valence electrons. 
For Mn$_2$RuGa, $N_v$=25, resulting in a net moment of \SI{+1}{\bohrmagnetron}.
Mn$_2$Ga was expected to have a half-Heusler ($C1_b$) structure,\cite{PhysRevB.66.134428} with Mn on $4a$, $4c$ sites and Ga on $4b$ sites, thus leaving the $4d$ site empty.
In half-metallic Heuslers with the $C1_b$ structure, the magnetic moment $m$ is given by $m=N_v-18$. 
For Mn$_2$Ga, $N_v=17$ and hence $m=\SI{-1}{\bohrmagnetron}$. 
The idea behind the Mn$_2$Ru$_x$Ga system, as proposed by Kurt et al.\cite{coey2014}, 
was that by changing the Ru content $x$ a material can be formed mid-way between Mn$_2$RuGa and Mn$_2$Ga, 
that is half-metallic, yet presents no net magnetic moment.  

Cubic Mn$_2$Ru$_x$Ga was stabilized in thin film form by Kurt et al.\cite{coey2014},
who showed that at $x \approx 0.5$ there is an ordered spin state with a critical temperature of approximately $\SI{550}{\kelvin}$ and a very small net magnetic moment.
In the same work Andr\'eev reflection (PCAR) spectroscopy showed a \SI{54}{\percent} spin polarization at the Fermi level.
This is less than the \SI{100}{\percent} expected for an ideal half metal, 
but close to the values measured in other Heusler half-metals in thin film form.
Further evidence of half-metallicity comes from the spontaneous Hall angle, more than an order of magnitude higher than that observed in the $3d$-transition metals. 
This and the linear variation of $m$ with $x$ provided strong indications that Mn$_2$Ru$_{0.5}$Ga, grown by Kurt et al. \cite{coey2014} was a CFHM.

However, despite the experimental evidence,
the measurements do not match the theoretical understanding of Mn$_2$Ru$_x$Ga
provided by Galanakis et al.\cite{blugel2014}.
There are three main areas where disagreement between theory and experiment exists, namely 1) the on-set of half-metallicity; 2) the cell volume; and 3) the dependence of $M$ on $x$.

\begin{figure}
 \includegraphics[width=0.4\textwidth]{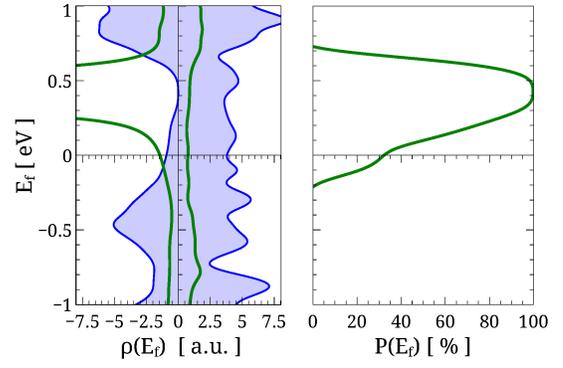}
	\caption{\label{dos} a) Spin resolved resistivity (green) and the corresponding density of states (DOS) (blue) calculated for the 
	Mn$_2$Ru$_{0.5}$Ga. The data corresponding to the spin down channel is represented with negative values.
      b) Calculated transport spin polarization.}
\end{figure}

\paragraph{Half-metallicity}
Density functional theory (DFT) investigations of Mn$_2$Ru$_{0.5}$Ga indicate that, although it may be possible to engineer a phase 
with zero magnetic moment, this will not be half-metallic~\cite{fesler2006,blugel2014,yang}.
Our calculations and those of others~\cite{blugel2014} 
find that the spin gap in the density of states (DOS) of Mn$_2$Ru$_{0.5}$Ga lies about \SI{0.4}{\electronvolt} above the Fermi level, $E_\mathrm{F}$. 
In Fig.~\ref{dos} we plot the calculated DOS and the corresponding resistivity, obtained by solving the Boltzmann equations in the relaxation time 
approximation. Although the DOS exhibits a large spin polarisation, $\approx \SI{60}{\percent}$, this is not reflected in that of the resistivity, which is 
only \SI{30}{\percent}. 
Since our transport calculations are expected to overestimate the spin polarization, the presence of the 'pseudo-gap'\cite{pseudogap} cannot justify 
the large experimentally observed spin polarization of the current.

\paragraph{Cell volume}
X-ray diffraction (XRD) and transmission electron microscopy (TEM) demonstrate that the films grow epitaxially with the in-plane lattice parameter 
being dictated by the MgO substrate ($a_{\text{MRG}} = \sqrt{2} a_{\text{MgO}})$ for all film thicknesses grown, while the out-of-plane 
lattice parameter $c$, depends strongly on the film thickness~\cite{coey2015}. This results in cell volumes much larger than those predicted by DFT;
and their variation with film thickness cannot be explained.

\paragraph{Magnetism}
The measured magnetic moment as a function of Ru concentration is linear\cite{coey2014} for $0.3 \leq x \leq 0.7$ with slope $d M / d x = 2$, 
suggesting that there is a spin gap at the Fermi level over an extended range of concentrations, in contradiction to the DFT results.


\begin{table}
\begin{ruledtabular}
\begin{tabular}{cccccccccc}
  \# & $4a$ & $4c$ & $4b$ & $4d$  & $\Delta H$ [\SI{}{\electronvolt}] & M [\SI{}{\bohrmagnetron}] & $c/a$ & Vol [\SI{}{\angstrom}$^3$] \\ \hline\noalign{\smallskip}
1& Mn & Ru & Ga & Ru  & -1.11 & 2.18 & 1.0 & 56.33\\
2 & Ga & Mn & Ga & Ru  & -0.99 & 2.93 & 1.0 & 55.83\\
3& Mn & Mn & Ga & Ru  & -0.97 & 0.07 \footnote{Magnetization quenched by the tetragonal distortion.} & 1.2 & 55.88\\
4& Mn & Mn & Ga & Ru  & -0.52 & 4.66 \footnote{Ferromagnetic Mn$_2$RuGa phase.} & 1.0& 55.01\\
\vdots & \vdots & \vdots & \vdots & \vdots & \vdots & \vdots & \vdots & \vdots \\
10 & Ga & Mn & Ga & --  & 0.02 & 3.14 & 1.0 & 52.01\\
12& Ru & Mn & Ga & --  & 0.27 & 0.19 & 1.0 & 44.70\\
13& Ru & Mn & Ga & Ru  & 0.29 & 4.44 & 1.0 & 59.20\\
14& -- & Mn & Ga & Ru  & 0.52 & 4.50 & 1.0 & 49.60\\
15 & Mn & Mn & Ga & --  & 0.54 & 0.47 & 1.0 & 46.54\\
\vdots & \vdots & \vdots & \vdots & \vdots & \vdots & \vdots & \vdots & \vdots \\
\end{tabular}
\end{ruledtabular}
\caption{Calculated enthalpies of formation, $\Delta H$, for the most stable competing Heusler phases of the 1221 structural and magnetic 
  Mn-Ru-Ga cells investigated. The configurations investigated are limited to the primitive 4-atom Heusler cell (3-atom for half-Heuslers).\cite{mm}
}\label{table}
\end{table}
Here we address and resolve the conflict between experiment and theory.
By applying a high-throughput approach based on the \text{VASP}\cite{vasp} 
implementation of DFT and the Perdew-Burke-Ernzerhof (PBE)\cite{PBE} functional,
we have calculated the properties of $1221$ Heusler phases 
containing Mn, Ru and Ga, that present different stoichiometry,
magnetic order and site occupancy within the L2$_1$ and C1$_b$ symmetries. 
Other symmetries were excluded from the search. 
For each configuration we compute the enthalpy of formation, $\Delta H$,
with respect to the lowest energy phase of each of the constituent elements,
allowing us to compare the relative stability of the various configurations.
Our results for the lowest energy structures are summarized in Table~\ref{table}. 

For the Mn$_2$RuGa composition,
we find that the lowest energy structure corresponds to Mn occupying the inequivalent $4a$,
$4c$ sites, Ga the $4b$ and Ru taking the remaining $4d$ site,
consistent with literature\cite{coey2014,hori}.
Cubic Mn$_2$Ga was found to have a positive enthalpy of formation of \SI{0.54}{\electronvolt \per \formulaunit},
making the compound unstable with respect to decomposition into its elementary phases.
The symmetry of the stable D0$_{22}$ structure was excluded from our calculations. 
However, the most energetically favourable Mn$_2$Ga structure in the L2$_1$ phase (\# 15), places Mn on the inequivalent $4a$ and $4c$ sites with Ga occupying the $4b$ one.
The structure remains cubic and the magnetic state is ferrimagnetic,
consistent with experimental characterisation presented by Kurt ~et al.~\cite{coey2014} 
for thin films stabilized on a suitable substrate or a seed layer.
We note that the formation of any half-Heusler in the Mn, Ru and Ga phase diagram is energetically unfavourable, 
so that we would not expect pure half-Heuslers to be a significant constituent of the films.

\begin{figure}
\includegraphics[width=0.4\textwidth]{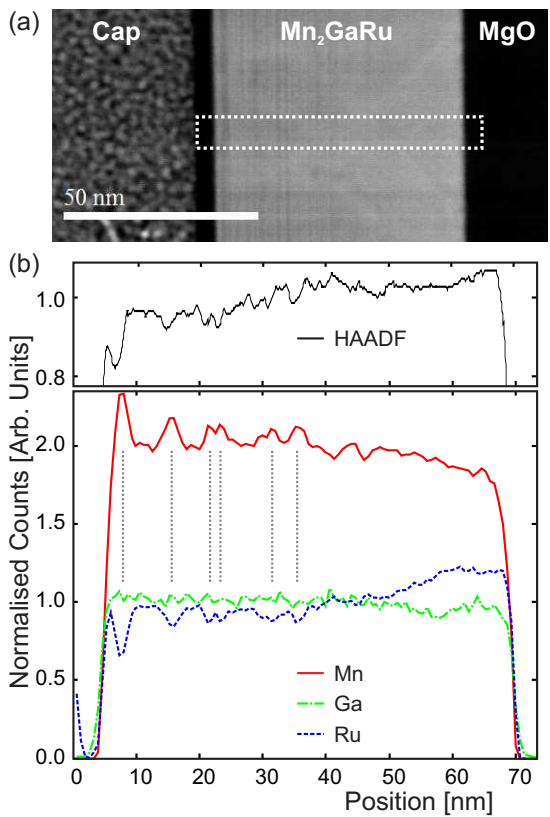}
\caption{\label{fig:tem} 
Analytical electron microscopy analysis of the thin film ($x = 1$) composition.
(a) High angle annular dark field (HAADF) image of a typical sample cross-section, 
with the region where EELS data were acquired indicated by a white rectangle. 
In HAADF images, heavy elements appear brightest.
(b) Elemental variations across the thin film stack, as measured by EELS. 
Data are normalised by setting the average composition to Mn$_2$GaRu.
The dashed vertical lines indicate regions where of Mn enrichment and Ru depletion, corresponding to darker features in the HAADF image.}
\end{figure}
We note that the energy of the system can be lowered by forming Mn-deficient MnRu$_2$Ga or MnGa$_2$Ru (\# 1 and 2 in table ~\ref{table}). 
Given that the difference in formation enthalpy of these phases with respect to Mn$_2$RuGa (\# 3 in table~\ref{table}) is only 
\SI{0.14}{\electronvolt\per\formulaunit}, we expect that actual samples will not form distinct polycrystalline 
phases, but instead display significant site disorder, particularly on the $4a$ site, with a preference towards a lower Mn content.
This was confirmed by laser-assisted inductively coupled mass spectroscopy (ICPMS) measurements of the Mn-to-Ga ratio for a series of 
samples with varying Ru concentration, $x$. 
The ratio was observed to be in the \num{1.6}-\num{1.9} range, increasing with increasing film 
thickness.

In Fig. \ref{fig:tem} (a) we show scanning transmission electron microscopy (STEM) measurements of electron-transparent lamell{\ae} of Mn$_2$RuGa
which indicate that there is little variation in either the in-plane or out-of-plane lattice constants throughout the film. 
The corresponding electron energy loss (EELS) spectra and line profiles are shown in Fig. \ref{fig:tem} (b).  
Alternating light and dark bands in dark-field images indicate slight compositional variations, especially in the layers closest to the surface. 
EELS measurements reveal that these bands correspond to layers of Mn enrichment and Ru depletion, 
suggesting a degree of phase segregation during growth, but not at the level of formation of half-Heuslers. 
We also note that the Ru concentration, $x$, decreases by about \SI{20}{\percent} from the interface with the substrate through the thickness of the film; 
to a lesser extent, the Mn concentration increases across the same range.

In order to investigate the properties of low-Mn-content films we have performed supercell calculations where $1/3$ of the Mn atoms at the $4a$ site 
are substituted with Ga. The Mn-Ga substitution simultaneously changes the lattice parameters, the magnetic properties and the electronic 
structure of the system. We find that the ionic charges of Mn and Ga are $+2$ and $+1$, respectively. Hence a one-atom Mn-Ga substitution 
leaves the system with one unbound electron, thus creating electronic doping. 
Below we describe in detail the properties of such Mn-deficient
compounds.
\begin{figure}
  \includegraphics[width=0.8\columnwidth, clip=true, trim =4 0 3 0] {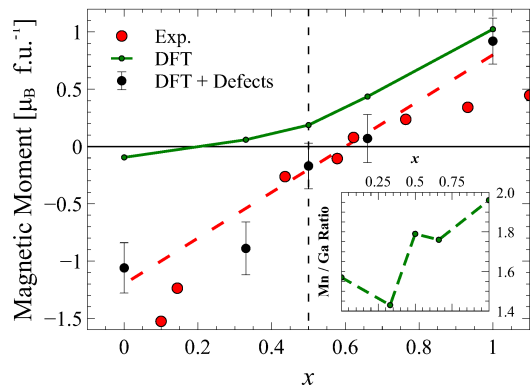}
	\caption{\label{fig_magnetization} Comparison of the magnetic moment predicted by DFT with the experimental results.
	The calculated magnetic moment for ideal MRG compounds is shown by the green line. Correction to the DFT magnetization 
	due to the formation of Ga defects is shown by the black points. The corresponding estimate of the Mn/Ga ratio is shown in the inset.
	The dashed line equivalent to doping of $\SI{2}{\elementarycharge\per Ru}$ is shown to guide the eye.
}
\end{figure}

\begin{figure}
\includegraphics[width=0.4\textwidth]{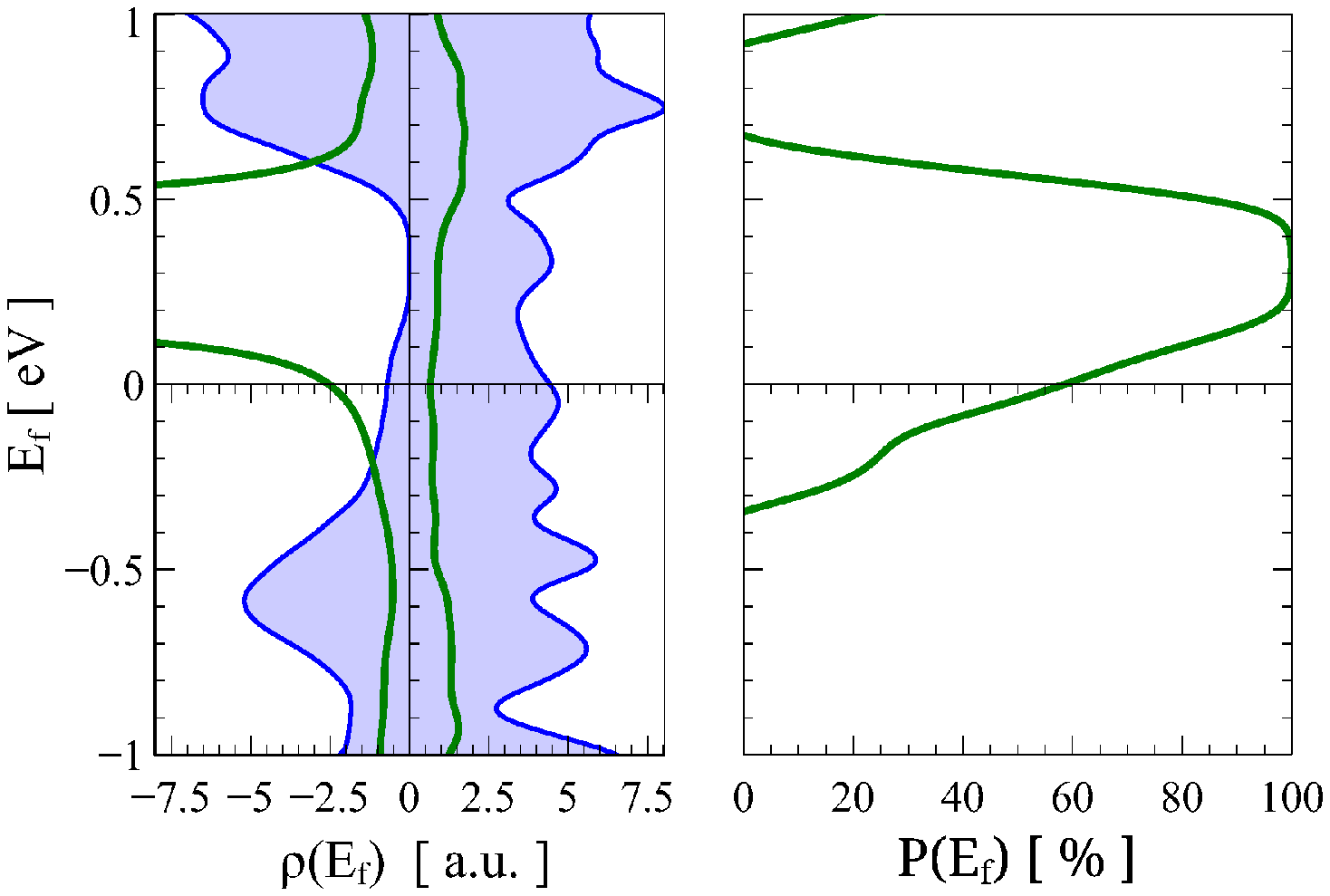}
\caption{\label{doped_dos} 
a) Spin resolved resistivity (green line) and the corresponding DOS (blue shadowed area) for electron-doped Mn$_2$Ru$_{0.5}$Ga. 
The doping was fixed at $0.4$ electrons per formula. The data corresponding to the spin down channel are represented by negative values. 
b) Calculated transport spin polarization.}
\end{figure}

\paragraph{Lattice}
Electronic doping provides an explanation for the variation of the lattice parameter with film thickness. From the volume difference 
between the relaxed DFT structure and the corresponding experimental one and by using the bulk modulus $B_0$, we estimate the 
experimental electronic doping. The bulk modulus is calculated for Mn$_2$Ru$_x$Ga ($x=~$\numlist{0.0; 0.33; 0.50; 0.66; 1.0}) compounds by 
fitting the Murnaghan equation of state\cite{EOS, murnaghan1944}.
By using a simple model 
\begin{equation}
\label{eq_doping}
n_\mathrm{el} = \frac{B_0}{S_0} \left[\left(\frac{c}{a}\right)_\mathrm{exp} \cdot \left(\frac{a_\mathrm{exp}}{a_0}\right)^3 - 1 \right]~\:,
\end{equation}
we can relate the experimentally observed lattice parameters to the electron doping level,
$n_\mathrm{el}$. In Eq.(\ref{eq_doping}) $S_0$ is the rate of 
change of the excess pressure with electron doping,
while $a_\mathrm{exp}$ and $a_0$ correspond to the experimental in-plane lattice constant
and the relaxed theoretical lattice constant, respectively.
This equation is easily derived under the assumption that the material is in mechanical equilibrium 
at the experimental lattice constant,
due to the excess pressure provided by the Mn-Ga substitution,
via the electron doping mechanism.
Since we are comparing pressure differences, we ignore constant pressure terms.
In order to stabilise the experimental lattice parameters,
including the observed $c/a > 1$,
we find electron doping in the range \SIrange{0.1}{0.5}{\elementarycharge\per\formulaunit}, 
corresponding to a Mn/Ga ratio in the interval 1.4 to 2.0.
The higher doping level occurs for the lower Ru concentrations,
as shown in the inset of Fig.~\ref{fig_magnetization}.

TEM imaging and spectroscopy clearly indicate that there are regions of high Mn content,
and therefore regions of enhanced Ga content elsewhere in the sample.
It is therefore reasonable to assume that films of different thicknesses will have a different electronic doping,
which in turn alters the $c$-lattice parameter and does so uniformly throughout the sample. 

\paragraph{Magnetism}
A key feature of a CFHM is the magnetically compensated ground state.
In agreement with Ref.\ \cite{blugel2014},
we find that the magnetization calculated for MRG compounds,
as a function of the Ru doping, 
differs from the experimental one (see Fig. \ref{fig_magnetization}). 
The discrepancy is two-fold:
I) the slope of the magnetization with $x$ disagrees by a factor of 2, and 
II) there is no compensation of the magnetization around $x=0.5$. 
These discrepancies are resolved if we take into account the effect of the Mn-Ga substitution on the magnetic properties.  
Ga defects introduce, in addition to the electronic doping,
a change in the net magnetic moment per unit cell, 
of \SI{-2}{\bohrmagnetron} per Mn substituted by Ga,
which allows us to express the expected moment $M_\mathrm{EXP}$ as
\begin{equation}
\label{eq_magnetization}
M_\mathrm{EXP}(x) = M_\mathrm{DFT}(x) - 2 \cdot n_\mathrm{el}(x)\:,
\end{equation}
where $M_\mathrm{DFT}$ is the theoretically calculated magnetic moment for a defect-free Mn$_2$Ru$_x$Ga compound.

The corrections given by Eqns.~(\ref{eq_doping}) and (\ref{eq_magnetization})
have been applied for each value of $x$,
and the results are summarised in Fig.~\ref{fig_magnetization}.
Notably, the presence of defects improves significantly the agreement between the 
experimental and theoretical magnetic moments,
with the exception of concentrations around $x=1$.
A neutron diffraction study by Hori et al.~\cite{hori} has shown that the magnetization
of stoichiometric Mn$_2$RuGa ($x=1$) is $\approx\SI{1}{\bohrmagnetron}$, 
in good agreement with our calculations.
This leads us to the conclusion that in the $x=1$ limit there may be a substantial content of 
the Ru$_2$MnGa phase, which is known to be antiferromagnetic~\cite{hori}.

\paragraph{Electronic Structure}
Finally, we discuss the effect of the electronic doping on the degree of transport spin polarisation.
In figure~\ref{doped_dos} we show the DOS and corresponding Boltzmann resistivity
for Mn$_2$RuGa with $n_\mathrm{el}=0.4$ extra electrons per formula unit,
coresponding to a Mn/Ga ratio of $1.6$ as observed by ICPMS.
The additional doping results in a transport spin polarization of \SI{\approx 60}{\percent},
which is twice as large as the one calculated for the original 
Mn$_2$Ru$_x$Ga compound.
At a doping level of $n_\mathrm{el} = 1.0$ the transport spin polarization becomes \SI{100}{\percent}.
It is important to note that the calculations presented here
do not take into account the effect of the disorder due to the Mn-Ga substitution
on the transport properties.
We anticipate that the presence of disorder may open further the spin gap,
resulting in a cumulative effect where the disorder provides both the spin gap and
the electron doping necessary for reinstating the half-metallicity.
The improvement of spin-transport properties obtained by introducing disorder,
has already been discussed by Chadov et al.~\cite{chadov2012}.

In conclusion, a sustained dialogue between experimental measurements and theoretical calculations have demonstrated 
that Mn$_2$Ru$_x$Ga can form a true CFHM.
As a consequence we expect that it will become a cornerstone for future spintronics 
technology.
By means of high-throughput calculations, we have shown 
that there are several competing phases in the Mn-Ru-Ga system;
and that due to their small energy differences
they exhibit a strong tendency towards site disorder, and a preference for reduced Mn content.
This has all been confirmed by our experimental characterization of MRG thin films.
Furthermore the low Mn content provides an electronic doping mechanism,
pushing the system towards half-metalicity and
improving the agreement between experiment and theory regarding the structural and magnetic properties of the system. 
Based on our calculations, complete transport spin polarisation can be achieved.

We have shown that chemical composition, $c/a$ ratio,
tendency to site-disorder and cell volume are all correlated.
To achieve transport half-metallicity and zero net moment,
a reduced Mn to Ga ratio of $\approx 1.4$ is required,
as well as a Ru concentration of $\approx 0.7$.
Fine-tuning of the position of the Fermi level in the spin gap
can then be achieved through varying the $c/a$ ratio,
which we have shown can be done by varying the film thickness.

\begin{acknowledgments}
The authors would like to thank Cora McKenna for the ICPMS mesaurements,
and H. Kurt for fruitful discussions.
Computational resources have been provided by the Trinity Center for High Performance Computing (TCHPC) 
and by the Irish Centre for High-End Computing (ICHEC).
The authors acknowledge financial support from the FP7 project 'ROMEO' (grant number 309729),
Science Foundation Ireland through 'AMBER', and from grant 13/ERC/12561.
\end{acknowledgments}

\bibliography{nourl.bib}

\end{document}